\documentclass[pra,twocolumn,showpacs,preprintnumbers,amsmath,amssymb]{revtex4}
\usepackage{mathrsfs}
\usepackage{bbm}
\usepackage{amsfonts}
\usepackage{tipa}

\usepackage{epsfig,graphicx}
\usepackage{amstext}
\usepackage{amsmath}
\usepackage{graphicx}
\usepackage{times}

\begin{document}


\title{Sudden change of geometric quantum discord in finite temperature reservoirs}

\author{Ming-Liang Hu}
\email{mingliang0301@163.com}
\author{Jian Sun}
\affiliation{School of Science, Xi'an University of Posts and Telecommunications,
             Xi'an 710061, China}

\begin{abstract}
We investigate sudden change (SC) behaviors of the distance-based
measures of geometric quantum discords (GQDs) for two
non-interacting qubits subject to the two-sided and the one-sided
thermal reservoirs. We found that the GQDs defined by different
distances exhibit different SCs, and thus the SCs are the combined
result of the chosen discord measure and the property of a state. We
also found that the thermal reservoir may generate states having
different orderings related to different GQDs. These inherent
differences of the GQDs reveal that they are incompatible in
characterizing quantum correlations both quantitatively and
qualitatively.

\end{abstract}


\maketitle

\section{Introduction}\label{sec:1}
Due to the emergence of quantum information science, the
characterization and quantification of quantum correlations in a
system have became one of the people's research focuses in the past
few decades \cite{nielsen-qcqi}. Nowadays, when we mention to
quantum correlations, two prominent lines of research on this
problem may immediately appear in our minds. The first one centers
around the entanglement-separability paradigm, under which various
forms of the entanglement measures have been proposed and
extensively studied \cite{rmp-en}. The second one is based on the
noncommutativity of operators in quantum mechanics, and along this
line people also presented a plenty of discord-like quantum
correlation measures \cite{rmp-qd}.

It is well accepted that entanglement is responsible for the
advantage of many quantum communication and computation tasks
\cite{Ekert,DiVincenzo,Pan}. It is also realized recently that
quantum discord \cite{qd}, which reveals quantum correlations from a
different perspective, plays a vital role in quantum protocols such
as the deterministic quantum computation with one qubit \cite{DQC1},
remote state preparation \cite{rsp}, quantum locking of classical
correlations \cite{qlock1,qlock2}, quantum state broadcasting
\cite{qsb}, and quantum state merging \cite{qsm1,qsm2}. Moreover,
the discord consumption has been linked to the quantum advantage for
extracting information via coherent interactions \cite{Gumile}, and
meanwhile, it can also be connected to entanglement via some
measurement processes \cite{qd-en}.

Although both can serve as physical resources for quantum
information processing, quantum discord and entanglement are in fact
fundamentally different in many aspects. The most prominent one
is that quantum discord may be increased by local operations
\cite{qd-in}, while entanglement can only be increased by coherent
operations. Moreover, when considering their evolution for open
quantum system, quantum discord and entanglement also exhibit
distinct singular behaviors. For instance, quantum discord is more
robust against decoherence than entanglement
\cite{robu1,robu2,robu3}, it is immune to certain quantum noises
during certain time intervals \cite{froz1,froz2,froz3,add3,add4},
and therefore exhibits the frozen behavior which is impossible for
entanglement. Quantum discord exhibits sudden changes (SCs) due to
the optimization procedure involved in its definition
\cite{sc1,sc2,sc3,add1,add2,sc4,sc5}, while entanglement undergoes
sudden death \cite{ESD} which does not exist for quantum discord. It
is also worthwhile to note that the sudden death of entanglement is
independent of the chosen entanglement measures, as it occurs
whenever the evolved state becomes separable.

We investigate in this work the singular behaviors of three kinds of
the geometric quantum discords (GQDs) for two non-interacting qubits
subjecting to the independent thermal reservoirs. We will show that
for both the two-sided and the one-sided reservoirs, there are SCs
being observed during the evolution of the GQDs, and the critical
times for their occurrence are strongly dependent on the choice of
the discord measures, which means that this phenomenon is the
combined result of the chosen discord measure and the quantum state
other than a property of the state itself. This statement is further
confirmed by the relativity of different GQDs, i.e., different GQDs
may impose different orderings of quantum states.

We arrange this paper as follows. In Sections \ref{sec:2} and
\ref{sec:3}, we recall the formula for the three GQDs, and present
the analytical solution for the master equation describing the
evolution of the system. Then in Section \ref{sec:4}, we give a
discussion of the SC phenomenon. Finally, Section \ref{sec:5} is
devoted to a summary.

\section{Distanced-based measures of the GQDs}\label{sec:2}
As we mentioned in Section \ref{sec:1}, the quantum discord in a
system can be quantified from different perspectives. We adopt in
this work three forms of the distance-based measures of GQDs
\cite{trace1,trace2,trace-dyn,trace-ana,hellinger,lqu,bures1,bures2,bures-ana}.
They are defined via the minimal distance from $\rho$ to $\Omega_0$,
i.e.,
\begin{eqnarray}\label{eq1}
 D_\alpha(\rho)=\min_{\chi\in \Omega_0} d_\alpha(\rho,\chi)
\end{eqnarray}
where $d_\alpha$ designates the different distance measures one used
in defining the GQDs. They are all well defined and can avoid the
problem for the GQD based on the Frobenius norm \cite{gqd-fro}. We
concentrate in the following on a bipartite system $AB$ described by
the density operator $\rho$, and the three GQDs are defined
respectively via the trace distance
\cite{trace1,trace2,trace-dyn,trace-ana}, the Hellinger distance
\cite{hellinger,lqu}, and the Bures distance
\cite{bures1,bures2,bures-ana}. To facilitate later description, we
call them the TDD, HDD, and BDD for brevity.

We first recall the corresponding measures for the trace distance,
the Hellinger distance, and the Bures distance between $\rho$ and
$\chi$, which are given respectively by
\begin{equation}\label{eq2}
\begin{split}
 & d_{\rm T}(\rho,\chi)=\parallel\rho-\chi\parallel_1,\\
 & d_{\rm H}(\rho,\chi)=\parallel\sqrt{\rho} -\sqrt{\chi}\parallel_2,\\
 & d_{\rm B}(\rho,\chi)=[2(1-\sqrt{F(\rho,\chi)})]^{1/2},
\end{split}
\end{equation}
where $\parallel\cdot\parallel_1$ and $\parallel\cdot\parallel_2$
denote respectively the trace norm and the Frobenius norm, while
$F(\rho,\chi)= [{\rm tr}(\sqrt{\rho}\chi\sqrt{\rho})^{1/2}]^2$
represents the Uhlmann fidelity.

Then, the TDD, HDD, and BDD can be defined respectively via the
minimal $d_{\rm T}(\rho,\chi)$, $d_{\rm H}(\rho,\chi)$, and $d_{\rm
B}(\rho,\chi)$ as
\begin{equation}\label{eq3}
\begin{split}
 &D_{\rm T}(\rho)=\min_{\chi\in\Omega_0}\parallel\rho-\chi\parallel_1, \\
 &D_{\rm H}(\rho)=2\min_{\sqrt{\chi}\in\Omega'_0}\parallel\sqrt{\rho}-\sqrt{\chi}\parallel_2^2, \\
 &D_{\rm B}(\rho)=\sqrt{(2+\sqrt{2})\min_{\chi\in\Omega_0}(1-\sqrt{F(\rho,\chi)})},
\end{split}
\end{equation}
where the set $\Omega_0$ is usually taken to be the
classical-quantum state $\Omega_0=\sum_k (\Pi_k^A\otimes
I_B)\rho(\Pi_k^A\otimes I_B)$, and $\Omega'_0 =\sum_k
(\Pi_k^A\otimes I_B)\sqrt{\rho}(\Pi_k^A\otimes I_B)$, with
$\{\Pi_k^A\}$ being the projection-valued measurements. Moreover,
the constants $2$ and $2+\sqrt{2}$ before min are introduced for the
normalization of $D_{\rm H}(\rho)$ and $D_{\rm B}(\rho)$ for the
two-qubit maximally discordant states.

The calculation of the TDD, HDD, and BDD is a hard task for general
$\rho$, and analytical results exist only for certain special
classes of states. First, for the two-qubit {\it X} state $\rho^X$
which only contains nonzero elements along the main diagonal and
anti-diagonal, the TDD is given by \cite{trace-ana}
\begin{eqnarray}\label{eq4}
 D_{\rm T}(\rho^X)=\sqrt{\frac{\xi_1^2\xi_{\rm max}-\xi_2^2\xi_{\rm min}}
                   {\xi_{\rm max}-\xi_{\rm min}+\xi_1^2-\xi_2^2}},
\end{eqnarray}
where $\xi_{1,2}=2(|\rho_{23}|\pm |\rho_{14}|)$,
$\xi_3=1-2(\rho_{22}+\rho_{33})$, $\xi_{\rm
max}=\max\{\xi_3^2,\xi_2^2+x_{A3}^2\}$, and $\xi_{\rm
min}=\min\{\xi_1^2,\xi_3^2\}$, with
$x_{A3}=2(\rho_{11}+\rho_{22})-1$.

Second, for the $2\times n$ dimensional state $\rho$, and the
decomposed $\sqrt{\rho}=\sum_{ij} c_{ij}X_i\otimes Y_j$, with
$\{X_i:i=0,1,2,3\}$ and $\{Y_j:j=0,1,\ldots,n^2-1\}$ constituting the
orthonormal operator bases for the Hilbert spaces $\mathcal {H}_A$
and $\mathcal {H}_B$, the HDD can be calculated as \cite{hellinger}
\begin{eqnarray}\label{eq5}
 D_{\rm H}(\rho)=2(1-||\bold{r}||_2^2-z_{\max}),
\end{eqnarray}
where $||\bold{r}||_2^2=\sum_j c_{0j}^2$, and $z_{\max}$ represents
the largest eigenvalue of the matrix $ZZ^\dag$, with $Z=(c_{ij})_{
i=1,2,3;j=0,1,\cdots,n^2-1}$.

Finally, although there is no analytic solution, the calculation for
the maximum of $F(\rho,\chi)$ can be simplified as \cite{bures-ana}
\begin{eqnarray}\label{eq6}
 F_{\max}(\rho,\chi)=\frac{1}{2}\max_{||\vec{u}=1||}
 \left(1-{\rm tr}\Lambda+2\sum_{k=1}^{n}\lambda_k(\Lambda)\right),
\end{eqnarray}
where $\lambda_k(\Lambda)$ denote the eigenvalues of
$\Lambda=\sqrt{\rho}(\vec{u}\cdot\vec{\sigma}\otimes
I_B) \sqrt{\rho}$ in non-increasing order, with
$\vec{u}$ being a unit vector in $\mathbb{R}^3$, and
$\vec{\sigma}=(\sigma^x,\sigma^y,\sigma^z)$ is the vector of the
Pauli operators.

From the above equations, one can note that even for the simple
\emph{X} state, there are optimization procedures involved for
obtaining the GQDs, and this may induce SC behaviors of $D_{\rm
\alpha}(\rho)$. We will discuss them explicitly in the following
text.

\section{Solutions of the model}\label{sec:3}
The central system we considered consists of two qubits (labeled as
$S=A,B$) with large enough spatial distance, and the direct
interaction between them can be ignored. We will discuss two
different cases: (i) both qubits $A$ and $B$ are embedded in their
own independent thermal reservoirs, and (ii) only qubit $A$ (or $B$)
is embedded in the thermal reservoir, while the other one is free of
noise.

For the qubit $S$ subject to the thermal reservoir, the evolution of
$\rho^S(t)$ is governed by the master equation \cite{master}
\begin{equation}\label{eq7}
 \frac{d\rho^S }{dt}=\frac{\gamma_S}{2}\sum_{k=1}^2 \left(2\mathcal {L}_k^S
                    \rho^S\mathcal {L}_k^{S\dag}
                   -\{{L}_k^{S\dag}\mathcal {L}_k^S,\rho^S\}\right),
\end{equation}
under the Markovian approximation. Here, $\gamma_S$ is the strength
of the damping rate, $\{\cdot,\cdot\}$ denotes the anticommutator,
while $\mathcal {L}_1^S =\sqrt{\bar{n}+1}\sigma_S^{-}$ and $\mathcal
{L}_2^S =\sqrt{\bar{n}}\sigma_S^{+}$ (with $\sigma_S^{\pm}$ being
the raising and the lowering operators) describe, respectively, the
decay and excitation processes of the qubit $S$, with rates depending on
the temperature which is proportional to the average thermal photons
$\bar{n}$ in the reservoir.

The master equation \eqref{eq7} can be solved analytically. For
convenience of later presentation, we define
\begin{eqnarray}\label{eq8}
  q_1^S=\frac{\bar{n}+(\bar{n}+1)p_S^2}{2\bar{n}+1},
  ~q_2^S=\frac{\bar{n}(1-p_S^2)}{2\bar{n}+1},
\end{eqnarray}
then one can obtain that $\rho^S(t)$ takes the following form
\begin{widetext}
\begin{eqnarray}\label{eq9}
\rho^S(t)=\left(\begin{array}{cc}
      q_1^S\rho^S_{11}(0)+q_2^S\rho^S_{00}(0)   &  p_S\rho^S_{10}(0) \\
      p_S\rho^S_{01}(0)  & 1-q_1^S\rho^S_{11}(0)-q_2^S\rho^S_{00}(0) \\
            \end{array}
\right),
\end{eqnarray}
\end{widetext}
where $\rho_{ij}^S=\langle i|\rho^S|j\rangle$ in the standard basis
$\{|1\rangle,|0\rangle\}$ expanded by the eigenvectors of the Pauli
operator $\sigma_S^z$, and the time-dependent factor
$p_S=e^{-(2\bar{n}+1)\gamma_S t/2}$. Moreover, when $\bar{n}=0$, the
reservoir is at zero temperature, and the solution of Eq.
\eqref{eq9} reduces to that given in Ref. \cite{Bellomo}.

From the analytical expression for the single-qubit reduced density
matrix $\rho^S(t)$, one can obtain the two-qubit density matrix
$\rho(t)$ for arbitrary initial state $\rho(0)$ by using the
procedure presented in Ref. \cite{Bellomo}. The diagonal elements
are given by
\begin{equation}\label{eq10}
\begin{split}
 &\rho_{11}(t)=q_1^A q_1^B \rho_{11}(0)+q_1^A q_2^B \rho_{22}(0)+q_2^A q_1^B \rho_{33}(0)\\
              &\hspace{13mm} +q_2^A q_2^B \rho_{44}(0),\\
 &\rho_{22}(t)=q_1^A [(1-q_1^B) \rho_{11}(0)+(1-q_2^B) \rho_{22}(0)]\\
              &\hspace{13mm} +q_2^A [(1-q_1^B) \rho_{33}(0)+ (1-q_2^B)\rho_{44}(0)],\\
 &\rho_{33}(t)=(1-q_1^A) [q_1^B \rho_{11}(0)+ q_2^B \rho_{22}(0)]\\
              &\hspace{13mm} +(1-q_2^A)[q_1^B \rho_{33}(0)+q_2^B \rho_{44}(0)],\\
 &\rho_{44}(t)=1-\rho_{11}(t)-\rho_{22}(t)-\rho_{33}(t),
\end{split}
\end{equation}
while the nondiagonal elements are given by
\begin{equation}\label{eq11}
\begin{split}
 &\rho_{12}(t)=q_1^A p_B \rho_{12}(0)+q_2^A p_B \rho_{34}(0),\\
 &\rho_{13}(t)=q_1^B p_A \rho_{13}(0)+q_2^B p_A \rho_{24}(0),\\
 &\rho_{14}(t)=p_A p_B \rho_{14}(0),~\rho_{23}(t)=p_A p_B \rho_{23}(0),\\
 &\rho_{24}(t)=p_A (1-q_1^B) \rho_{13}(0)+p_A (1-q_2^B) \rho_{24}(0),\\
 &\rho_{34}(t)=(1-q_1^A)p_B \rho_{12}(0)+(1-q_2^A)p_B\rho_{34}(0),
\end{split}
\end{equation}
and the other nondiagonal elements can be written directly by using
the Hermitian condition $\rho_{ij}(t)=\rho_{ji}^*(t)$.

For the special case of $p_A=p_B=p$, we obtain the two-sided
identical reservoir $\mathcal {E}_{AB}=\mathcal {E}_A\otimes\mathcal
{E}_B$, while for $p_A=p$ and $p_B=1$ ($p_A=1$ and $p_B=p$), it
reduces to the one-sided reservoir $\mathcal {E}_A$ ($\mathcal
{E}_B$).

\section{SC of the GQDs}\label{sec:4}
Based on the solutions in Eqs. \eqref{eq10} and \eqref{eq11} for the
system-reservoir coupling mode presented in Section \ref{sec:3}, we
begin to discuss decay dynamics of the three GQDs. We will show
that they exhibit distinct singular behaviors, which include the
completely different SCs and their relativity on
characterizing quantum correlations. To be explicit, we consider
initial two-qubit state of the following form
\begin{eqnarray}\label{eq12}
 |\Psi\rangle=\alpha|11\rangle + \beta|00\rangle,
\end{eqnarray}
where $\alpha\in[0,1]$, and $\beta=\sqrt{1-\alpha^2}$. The
analytical expressions of $\rho(t)$ for both the two-sided identical
reservoirs $\mathcal {E}_{AB}$ and the one-sided reservoir $\mathcal
{E}_A$ (or $\mathcal {E}_B$) can be written directly from Eqs.
\eqref{eq10} and \eqref{eq11}, which are of the \emph{X} form.

\begin{figure}
\centering
\resizebox{0.46\textwidth}{!}{%
\includegraphics{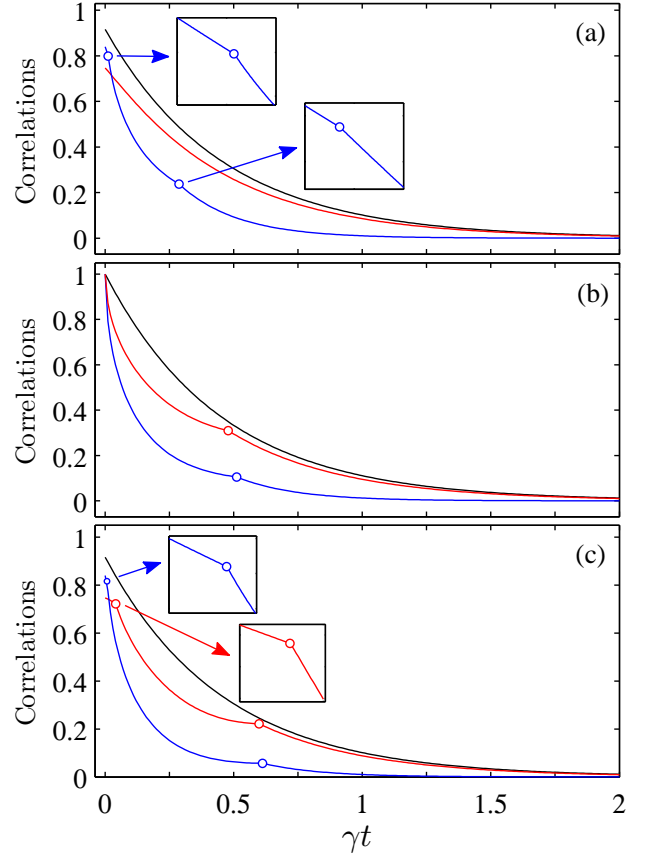}}
\caption{(Color online) $\gamma t$ dependence of $D_{\rm T}(\rho)$
(black), $D_{\rm B}(\rho)$ (red), and $D_{\rm H}(\rho)$ (blue) for
the initial state $|\Psi\rangle$ subject to the two-sided reservoir
$\mathcal {E}_{AB}$ with $\bar{n}=0.6$. The other parameters are
given by $\alpha^2=0.3$ (a), $0.5$ (b), and $0.7$ (c). The hollow
circles denote the SC points.} \label{fig:1}
\end{figure}

We first consider the two-sided identical reservoir $\mathcal
{E}_{AB}$. Fig. \ref{fig:1} is an exemplified plot of the $\gamma t$
dependence of $D_{\rm T}(\rho)$, $D_{\rm B}(\rho)$, and $D_{\rm
H}(\rho)$ for the initial state $|\Psi\rangle$ with $\bar{n}=0.6$
and different values of $\alpha^2$. For this case, as
$\rho_{23}(t)=0$, the TDD can be obtained analytically as
\begin{equation}\label{eq13}
 D_{\rm T}(\rho)=2p^2\alpha\sqrt{1-\alpha^2},
\end{equation}
therefore it is symmetric with respect to $\alpha^2=0.5$, and decays
smoothly and monotonously with increasing $\gamma t$ for any
$\alpha^2$.

The BDD and the HDD are no longer the symmetric functions of
$\alpha^2=0.5$. As displayed in Fig. \ref{fig:1}, while both $D_{\rm
B}(\rho)$ and $D_{\rm H}(\rho)$ still decay monotonously with
increasing $\gamma t$, there are also SCs being observed, i.e., they
are nonsmooth functions of $\gamma t$. In particular, the critical
time $\gamma t_c$ for the SCs and the times of SCs are determined
strongly by the chosen discord measure and the form of the initial state.
For the chosen parameters in Fig. \ref{fig:1}, $D_{\rm B}(\rho)$ decays smoothly
for $\alpha^2=0.3$, while $D_{\rm H}(\rho)$ experiences double SCs at
$\gamma t_c\simeq 0.0115$ and $0.287$, respectively. For $\alpha^2=0.5$, the
single SC of $D_{\rm B}(\rho)$ ($\gamma t_c\simeq 0.478$) occurs
earlier than that of $D_{\rm H}(\rho)$ ($\gamma t_c\simeq 0.5115$),
while for $\alpha^2=0.7$, both $D_{\rm B}(\rho)$ and $D_{\rm
H}(\rho)$ exhibit double SCs, where the first one of $D_{\rm
B}(\rho)$ ($\gamma t_c\simeq 0.0404$) occurs shortly after $D_{\rm
H}(\rho)$ ($\gamma t_c\simeq 0.0066$), and the second one of $D_{\rm
B}(\rho)$ ($\gamma t_c\simeq 0.5985$) turns out to be a little bit
earlier than that of $D_{\rm H}(\rho)$ ($\gamma t_c\simeq 0.6113$).

\begin{figure}
\centering
\resizebox{0.46\textwidth}{!}{%
\includegraphics{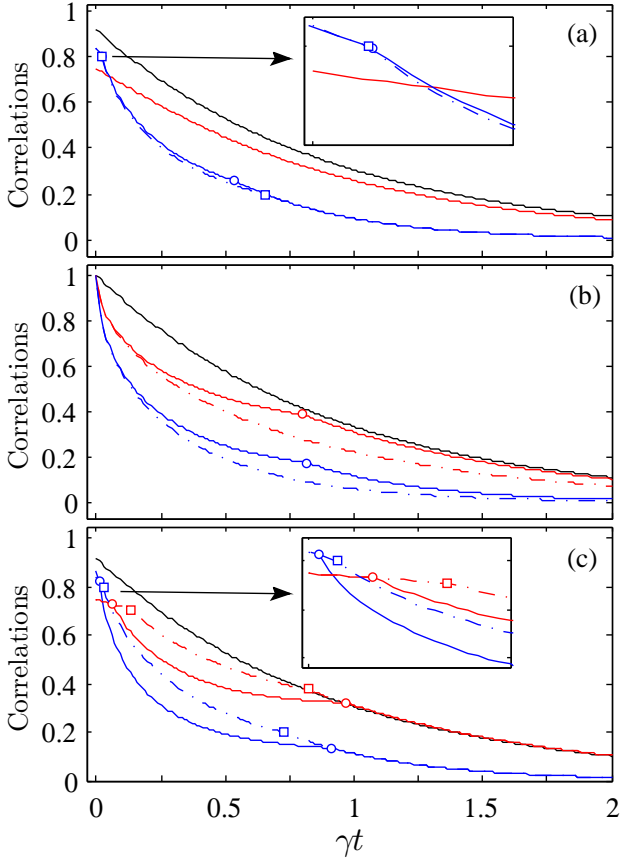}}
\caption{(Color online) $\gamma t$ dependence of $D_{\rm T}(\rho)$
(black), $D_{\rm B}(\rho)$ (red), and $D_{\rm H}(\rho)$ (blue) for
the initial state $|\Psi\rangle$ subject to the one-sided reservoir
$\mathcal {E}_{A}$ (solid) or $\mathcal {E}_{B}$ (dash-dotted) with
$\bar{n}=0.6$. The other parameters are $\alpha^2=0.3$ (a), $0.5$
(b), and $0.7$ (c). The hollow circles and squares denote the SC
points. Moreover, the lines of $D_{\rm T}(\rho)$, as well as the
lines of $D_{\rm B}(\rho)$ with $\alpha^2=0.3$, are overlapped for
the $\mathcal {E}_{A}$ and $\mathcal {E}_{B}$ cases.} \label{fig:2}
\end{figure}

Next we turn to discuss the cases of the one-sided reservoirs
$\mathcal {E}_{A}$ and $\mathcal {E}_{B}$. We displayed in Fig.
\ref{fig:2} the $\gamma t$ dependence of the GQDs by the solid and
the dash-dotted lines, respectively. As for these two cases, only
one of the two qubits is exposed to the reservoir, the decay of the
GQDs is slower than that for the two-sided reservoir case. First,
for both $\mathcal{E}_{A}$ and $\mathcal {E}_{B}$, the TDD is always
given by
\begin{eqnarray}\label{eq14}
 D_{\rm T}(\rho)=2p\alpha\sqrt{1-\alpha^2},
\end{eqnarray}
which is still a symmetric function about $\alpha^2=0.5$, and decays
smoothly and monotonously with the increasing $\gamma t$.

The BDD and the HDD may do not behave as smooth functions of $\gamma
t$. For the one-sided reservoir $\mathcal {E}_A$, they exhibit
qualitatively the same behaviors as those for the $\mathcal
{E}_{AB}$ case, and the only difference is that the critical times
for the SCs are all obviously delayed. For $\alpha^2=0.3$, the
double SCs for $D_{\rm H}(\rho)$ occur at $\gamma t_c\simeq 0.024$
and $0.536$, respectively. For $\alpha^2=0.5$, the single SC for
$D_{\rm B}(\rho)$ and $D_{\rm H}(\rho)$ occurs respectively at
$\gamma t_c\simeq 0.802$ and $0.816$. Finally, for $\alpha^2=0.7$,
the double SCs for $D_{\rm B}(\rho)$ occur at $\gamma t_c\simeq
0.0611$ and $0.966$, while for $D_{\rm H}(\rho)$ they occur at
$\gamma t_c\simeq 0.01045$ and $0.914$.

For the one-sided reservoir $\mathcal {E}_B$, although the evolved
density matrix differs only in $\rho_{22}(t)$ and $\rho_{33}(t)$
from the $\mathcal {E}_A$ case, the BDD and HDD are not exactly the
same (cf. the solid and the dash-dotted lines in Fig. \ref{fig:2}).
For $\alpha^2=0.3$, $D_{\rm B}(\rho)$ exhibits completely the same
$\gamma t$ dependence compared with that of the $\mathcal {E}_A$
case, while the double SCs of $D_{\rm H}(\rho)$ ($\gamma t_c\simeq
0.022$ and $0.652$) are slightly different. Moreover, the single SC
for both $D_{\rm B}(\rho)$ and $D_{\rm H}(\rho)$ disappears for
$\alpha^2=0.5$. Finally, for $\alpha^2=0.7$, the first SC of $D_{\rm
B}(\rho)$ ($\gamma t_c \simeq 0.1305$) and $D_{\rm H}(\rho)$
($\gamma t_c \simeq 0.027$) occurs later than that of the $\mathcal
{E}_A$ case, while their second SC ($\gamma t_c \simeq 0.822$ and
$0.724$, respectively) occurs earlier than that of the $\mathcal
{E}_A$ case.

The decay rates of the GQDs for $\mathcal {E}_A$ and $\mathcal
{E}_B$ may also be different. As showed in Fig. \ref{fig:2}, for
$\alpha^2 =0.5$, $D_{\rm B}(\rho)$ and $D_{\rm H}(\rho)$ for the
$\mathcal {E}_B$ case decay faster than those for the $\mathcal
{E}_A$ case in the whole $\gamma t$ region. For $\alpha^2 =0.3$ and
$0.7$, however, $\mathcal {E}_A$ and $\mathcal {E}_B$ give different
decay rates of $D_{\rm B}(\rho)$ and $D_{\rm H}(\rho)$ only during
limited $\gamma t$ regions. For $\alpha^2 =0.3$, $D_{\rm H}(\rho)$
for the $\mathcal {E}_B$ case decays slightly faster than that for
the $\mathcal {E}_A$ case when $\gamma t\in[0.022,0.652]$, while for
$\alpha^2 =0.7$, $D_{\rm B}(\rho)$ and $D_{\rm H}(\rho)$ for the
$\mathcal {E}_B$ case decay slower than those for the $\mathcal
{E}_A$ case when $\gamma t\in [0.0611,0.966]$ and $\gamma t \in
[0.01045,0.914]$, respectively.

\begin{figure}
\centering
\resizebox{0.46\textwidth}{!}{%
\includegraphics{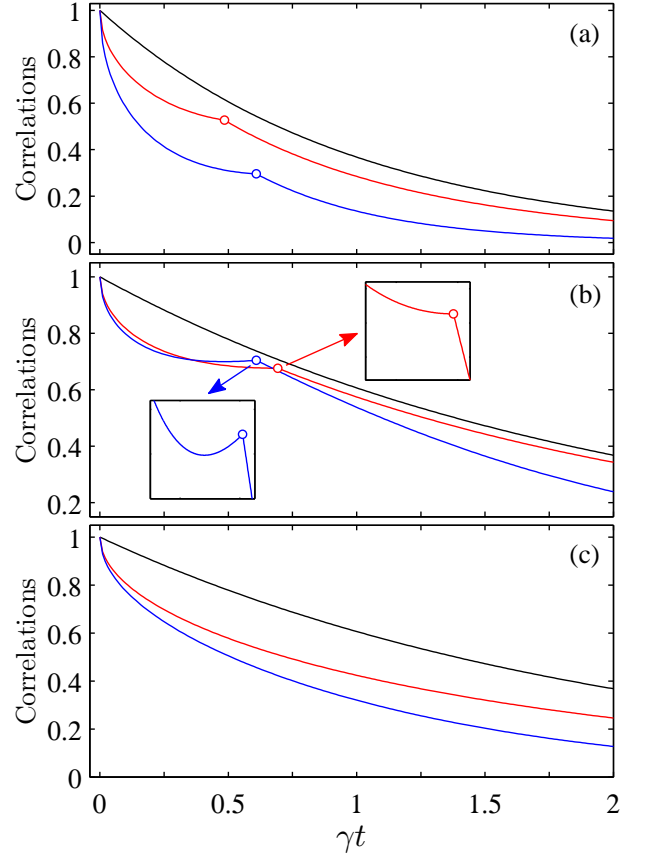}}
\caption{(Color online) $\gamma t$ dependence of $D_{\rm T}(\rho)$
(black), $D_{\rm B}(\rho)$ (red), and $D_{\rm H}(\rho)$ (blue) for
the initial state $|\Psi\rangle$ subject to the two-sided reservoirs
$\mathcal {E}_{AB}$ (a), one-sided reservoir $\mathcal {E}_{A}$ (b),
and $\mathcal {E}_{B}$ (c), all with the parameters $\bar{n}=0$ and
$\alpha^2=0.5$. The hollow circles denote the SC points.}
\label{fig:3}
\end{figure}

We discussed in the above evolution of the three GQDs, and observed
distinct singular behaviors such as the single and the double SCs
caused exclusively by the reservoir. We discuss in the following two
limiting cases, i.e., the zero temperature ($\bar{n}=0$) and the
infinite temperature ($\bar{n}\rightarrow \infty$) cases.

For the zero temperature case, $\mathcal {L}_2^S =0$, and the only
spontaneous decay term leads to a purely dissipative process. In the
long-time limit, this process drives the corresponding qubit to its
ground state $|1\rangle$, thus the GQDs disappear for arbitrary
initial state. In Fig. \ref{fig:3}, we showed the $\gamma t$
dependence of the three GQDs for $|\Psi\rangle$ with $\alpha^2=0.5$
and $\bar{n}=0$. For $\mathcal{E}_{AB}$, by comparing with Fig.
\ref{fig:1}(b), one can see that they exhibit very similar
behaviors, except that the decay rates are obviously decreased, and
the critical times for the SCs ($\gamma t_c \simeq 0.485$ and
$0.609$, respectively) are also slightly delayed. For
$\mathcal{E}_A$, there is also single SC for both $D_{\rm B}(\rho)$
($\gamma t_c \simeq 0.693$) and $D_{\rm H}(\rho)$ ($\gamma t_c
\simeq0.609$), which occurs earlier than those for the finite
temperature reservoir (cf. Fig. \ref{fig:3}(b) and Fig.
\ref{fig:2}(b)). Moreover, as showed by the blue line in Fig.
\ref{fig:3}(b), the HDD is increased with $\gamma t$ during the
region $\gamma t\in [0.482,0.609]$ (which is a reflection of the
fact that the quantum discord can increase under the local operation
on one party of the system \cite{nonmono}), while the TDD and the
BDD are always decreased. This means that different discord measures
may lead to different orderings of quantum states, and it confirmed
again that what the discord reveals is in fact the combined result
of the chosen discord and the quantum state other than a property of
the state itself. Of course, the increase for HDD is slight and
transient, and after $\gamma t> 0.609$, it decays to zero gradually.
Finally, for $\mathcal{E}_B$, the three GQDs still show
qualitatively the same $\gamma t$ dependence with those for the
finite temperature reservoirs, with however the decay rates are
evidently decreased.

For the infinite temperature case, we have $\mathcal {L}_1^S
=\mathcal {L}_2^S$, the decay and excitation processes occur at
exactly the same rate, and the noise induced by the transitions
between the two levels brings an arbitrary initial state into the
maximally mixed one. For concise of the paper, we do not present the
plots here. But the numerical results showed that for the initial
state $|\Psi\rangle$ with $\alpha^2=0.5$, all the three GQDs decay
smoothly and monotonously with the increasing $\gamma_0 t$ (
$\gamma_0=\bar{n}\gamma$), and there are no SCs being observed for
them.

Finally, as the SCs displayed in Figs. \ref{fig:1}, \ref{fig:2}, and
\ref{fig:3} are obtained via numerical methods, one may wonder
whether they are the real SCs or not, this is because sometimes it
is possible that what one observes as a SC might be the result of a
quick change that is actually not sudden when analyzed for smaller
time intervals \cite{add2}. For the TDD, as its analytical
expressions are given in Eqs. \eqref{eq13} and \eqref{eq14}, it is
evident that it does not experience SC for the model considered
here. For the HDD and BDD, as the square root of the density
operator $\rho$ cannot be derived analytically, analytical solutions
of $D_{\rm H}(\rho)$ and $D_{\rm B}(\rho)$ cannot be obtained. But
the changes observed in the three figures are actually exactly
sudden, and they are caused by the discontinuity of the optimal
angle related to measurement operators $\{\Pi_k^A\}$. In fact, by
writing $\Pi_{1,2}^A=(I_A\pm\vec{u} \cdot\vec{\sigma})/2$, and the
unit vector $\vec{u}=(\sin\theta\cos\phi, \sin\theta\sin\phi,
\cos\theta)$, we found that for the considered state in this paper,
both $D_{\rm H}(\rho)$ and $D_{\rm B}(\rho)$ are independent of the
angle $\phi$, but for the case of the single SC, the optimal
$\theta$ for both $D_{\rm H}(\rho)$ and $D_{\rm B}(\rho)$ is given
by $\pi/4$ before the SC point, and it changes abruptly to $\pi/2$
after the SC point. Moreover, for the case of double SCs, the
optimal $\theta$ changes abruptly from $\pi/2$ to $\pi/4$ and then
back to $\pi/2$ at the SC points. All these correspond to SCs of the
optimal $\{\Pi_k^A\}$, and therefore the changes observed for both
$D_{\rm H}(\rho)$ and $D_{\rm B}(\rho)$ are actually sudden.

\section{Summary}\label{sec:5}
In summary, we have investigated time evolution and the accompanying
singular behaviors of the TDD, BDD, and HDD. To focus exclusively on
the singular behaviors of them as caused solely by the thermal
reservoir, the two qubits of the central system are assumed to be
spatially separated far away from each other, and thus there are no
direct interactions between them, that is, every qubit interacts
with its own independent reservoir.

By solving analytically the master equation describing the evolution
of the two qubits, we analyzed dynamics of the three GQDs, and found
that they are incompatible in characterizing quantum correlations,
although they are all well defined from a geometric perspective. Our
findings are illustrated through two distinct behaviors of the three
GQDs. First, we found that the three GQDs may exhibit completely
different SCs for both the two-sided and the one-sided reservoirs.
The critical times for the SCs and the times of SCs are strongly
dependent on the choice of the GQD measure and the form of the
initial state. Different GQD measures have different SCs, and thus
the SCs are in fact the combined result of the chosen GQD measure
and the quantum state, but not the intrinsic property of a state
itself. This is fundamentally different from the sudden death of
entanglement, which is independent of the entanglement measure.
Moreover, we also revealed the relativity of different GQDs. To be
explicit, we found that the thermal reservoir may lead to a
generation of quantum states manifesting different orderings, and
this implies that different GQDs are incomparable as their behaviors
may not only be quantitatively but also be qualitatively different.

\section*{ACKNOWLEDGMENTS}
This work was supported by NSFC (11205121), and NSF of Shaanxi
Province (2014JM1008).

\newcommand{\PRL}{Phys. Rev. Lett. }
\newcommand{\RMP}{Rev. Mod. Phys. }
\newcommand{\PRA}{Phys. Rev. A }
\newcommand{\PRB}{Phys. Rev. B }
\newcommand{\PRE}{Phys. Rev. E }
\newcommand{\NJP}{New J. Phys. }
\newcommand{\JPA}{J. Phys. A }
\newcommand{\JPB}{J. Phys. B }
\newcommand{\PLA}{Phys. Lett. A }
\newcommand{\NP}{Nat. Phys. }
\newcommand{\NC}{Nat. Commun. }
%

%

\end{document}